\documentclass{article}

\usepackage[final,nonatbib]{my_nips_2017}

\usepackage[utf8]{inputenc} 
\usepackage[T1]{fontenc}    
\usepackage{hyperref}       
\usepackage{url}            
\usepackage{booktabs}       
\usepackage{amsfonts}       
\usepackage{nicefrac}       
\usepackage{microtype}      

\usepackage{mathtools}
\usepackage{subfig}

\usepackage{float}
\usepackage{epsfig}
\usepackage{amsmath}
\usepackage{amsthm}
\usepackage{amssymb}

\usepackage{algpseudocode}
\usepackage{framed}
\usepackage{longtable}
\usepackage{epstopdf}
\usepackage{url}
\usepackage{amsfonts}
\usepackage{amsmath}
\usepackage{MnSymbol}
\usepackage{mathrsfs}
\usepackage{graphicx}
\usepackage{caption}

\usepackage{wrapfig}

\numberwithin{equation}{section}

\newcommand{\eq}{\begin{equation*}}
\newcommand{\en}{\end{equation*}}
\newcommand{\eqa}{\begin{eqnarray*}}
\newcommand{\ena}{\end{eqnarray*}}
\newcommand{\eqn}{\begin{equation}}
\newcommand{\enn}{\end{equation}}
\newcommand{\be}{\begin{equation}}
\newcommand{\ee}{\end{equation}}
\newcommand{\eqan}{\begin{eqnarray}}
\newcommand{\enan}{\end{eqnarray}}

\newcommand{\M}{ {\bf M} }

\newcommand{\s}{ {\bf s} }

\newcommand{\y}{ {\bf y} }

\newcommand{\vv}{ {\bf v} }

\newcommand{\Dt}{ \Delta t }

\newcommand{\pmat}{\begin{pmatrix}}
\newcommand{\pman}{\end{pmatrix}}

\graphicspath{../figures/}
\graphicspath{{../figures/}}

\title{  Binary Bouncy Particle Sampler  }

\author{
  Ari Pakman 
 \\
Department of Statistics \\
Center for Theoretical Neuroscience \\
Grossman Center for the Statistics of Mind \\
Columbia University \\
\texttt{ari@stat.columbia.edu} 
}

\begin{document}

\maketitle

\begin{abstract}
The Bouncy Particle Sampler is a novel rejection-free non-reversible sampler for 
differentiable probability distributions over continuous variables.
We generalize the algorithm to piecewise differentiable distributions 
and apply it to  generic binary distributions using a piecewise differentiable augmentation.
We illustrate the new algorithm in a binary Markov Random Field example, and compare it to binary Hamiltonian Monte Carlo. 
Our results suggest that binary BPS samplers are better for easy to mix distributions. 
\end{abstract}

\section{Introduction}
The Bouncy Particle Sampler (BPS) algorithm is a novel generic sampler proposed in~\cite{peters2012rejection}
and explored in~\cite{monmarche2016piecewise,bouchard2017bouncy}.
Given a distribution $p(\y)$ with $\y \in \mathbb{R}^d$, 
the algorithm  introduces a random velocity vector $\vv$
distributed uniformly on the unit-sphere $\mathbb{S}^d$ and defines a piecewise deterministic Markov process~\cite{davis1984piecewise} over~$(\y, \vv)$. 
For reviews and further developments   see e.g.~\cite{fearnhead2016piecewise,pakman17SBPS,bierkens2017piecewise,vanetti2017piecewise}.
In this contribution we extend the basic BPS algorithm to 
piecewise differentiable distributions and apply it to generic binary discrete distributions using the augmentation method  of~\cite{pakman2013auxiliary}. 

\subsection{Discrete Infinitesimal Time Steps}
This section is a quick introduction to the BPS sampler for the 
reader unfamiliar with it, following closely the presentation in~\cite{pakman17SBPS}. 
We begin in discrete time and then take  the continuous-time limit. 
Let us introduce first the potential~$U(\y)$ as 
\eqan 
p(\y) \varpropto e^{-U(\y)}   \qquad \y \in \mathbb{R}^d
\label{puy}
\enan 
Denoting time by $t$, consider a discrete Markov process that acts on $(\y, \vv)$ as
\eqan 
(\y, \vv)_{t + \Delta t} = 
\left\{ \!\!\!\!\!
 \begin{array}{ll}
 (\y \! + \! \vv \Delta t  , \vv)   & \!\!\!\!\! \textrm{with prob. }   1-\Dt [\vv \cdot \nabla U(\y)]_+ 
\\
(\y \! + \! \vv \Delta t , \vv_r) &  \!\!\!\! \textrm{with prob. }   \Dt [\vv \cdot \nabla U(\y)]_+ 
\end{array}
\right.  
\label{pyse}
\enan 
where 
\eqan 
[x]_{+} = \max(x,0) \,,
\enan 
\eqan 
\vv_r =  \vv -2\frac{(\vv \cdot \nabla U(\y)) \nabla U(\y) }{||\nabla U(\y)||^2} \,.
\label{vvr}
\enan 
Note that 
 $\vv_r$ is  a reflection of $\vv$ with respect to  the plane perpendicular to the gradient $\nabla U$,
satisfying  $\vv_r \cdot \nabla U =-\vv \cdot \nabla U$ and $(\vv_r)_r = \vv$. 
In other words, the particle $\y$ moves along a straight line in the direction of $\vv$ and this direction is reflected as (\ref{vvr}) with probability $\Dt [\vv \cdot \nabla U(\y)]_+$.
This probability is non-zero only if the particle is moving in a direction of lower target probability $p(\y)$, or equivalently higher potential~$U(\y)$.

Remarkably, in the limit $\Delta t \rightarrow 0$,  the algorithm leaves the joint factorized distribution $p(\y)p(\vv)$ invariant. 
To see this, note that there are just two ways  to reach  $\y$ with velocity $\vv$ at time $t+\Delta t$. The first one is 
by being   at $\y -\vv\Delta t$ at time $t$ and moving a distance $\vv\Delta t$ without bouncing. This occurs with probability $1-\Delta t [\vv \cdot \nabla U]_+$.
The second possibility is that at time $t$ the particle was at $\y -\vv_r\Delta t$ with velocity $\vv_r$, moved $\vv_r\Delta t$ and bounced.  This event occurs with probability
$\Delta t [\vv_r \cdot \nabla U]_+ = \Delta t[-\vv \cdot \nabla U]_+$. Thus we have 
\eqan 
p_{t+ \Delta t}(\y, \vv)   &=&  (1-\Delta t [\vv \cdot \nabla U]_+ ) p_t(\y-\vv \Delta t)p_{t}(\vv)
+ \Delta t [-\vv \cdot \nabla U]_+ p_t(\y-\vv_r \Delta t)p_{t}(\vv_r)
\nonumber 
\\
&=& p_{t}(\vv) \left[   p_t(\y) - \Delta t   \vv \cdot \nabla p_t(\y) - \Delta t (\vv \cdot \nabla U) p_t(\y) \right] + O(\Delta t^2)
\label{pp}
\enan 
where we have used $p_{t}(\vv) = p_{t}(\vv_r)$ and
\eqan 
[\vv \cdot \nabla U]_+ - [-\vv \cdot \nabla U]_+ = \vv \cdot \nabla U
\enan 
Inserting now  (\ref{puy}), the second and third terms in (\ref{pp}) cancel and we get 
\eqan 
p_{t+ \Delta t}(\y, \vv) =    p_t(\y ) p_t(\vv ) +  O(\Delta t^2)
\enan 
which implies that the distribution is stationary, $\frac{d p_t(\y,\vv )}{dt} =0$.

\subsection{Continuous Time Limit for Integrable Distributions}
Applying the transition (\ref{pyse}) repeatedly and taking $\Delta t \rightarrow 0$, the random reflection point becomes 
an event in an inhomogeneous Poisson process with intensity  $[\vv \cdot \nabla U(\y)]_+$.
The resulting sampling procedure generates a piecewise linear Markov process~\cite{davis1984piecewise}.
To ensure ergodicity occasional resamplings are required in general~\cite{bouchard2017bouncy}, but not in the cases we will consider here.

The major challenge when applying the BPS algorithm is the sampling of
Poisson events with intensity  $[\vv \cdot \nabla U(\y)]_+$. In this work we  consider  distributions simple enough 
that this can be done with the inverse CDF method. In such cases, we initialize $(\y_0, \vv)$ and then 
iterate as many times as desired  the following steps:
\begin{enumerate}
 \item  Sample a uniform number $u \in [0,1]$
 
 \item   Move $\y$ in a straight line,
\eqan 
\y_{k} = \y_{k-1} + \vv t \,,
\enan 
where the time $t$ satisfies
\eqan 
u = e^{-\int_{0}^t \! dt' [\vv \cdot \nabla U(\y_{k-1} + \vv t')]_+ } \,.
\label{uet}
\enan 

\item Reflect the velocity as $\vv \rightarrow \vv_r$, defined in (\ref{vvr}).
\end{enumerate}

\section{Piecewise Continuous Distributions }
The algorithm described above can be easily extended to piecewise continuous distributions. 
Without loss of generality, assume $U(\y)$ is discontinuous across $y_1 =0$ 
and denote by $0^{\pm}$ the vector $\y$ in both sides of $y_1=0$, and by $t^{\pm}$
the time previous and posterior to the arrival to $y_1=0$.

The probability distribution is preserved if a particle that reaches $0^-$ with  $v_1>0$ at $t^-$, crosses to  
$0^+$ with probability 
\eqan 
q_{-+} = \min(1, e^{-U(0^+)+ U(0^-)  })\,,
\label{qpm}
\enan 
and, in case of rejection, inverts the sign of v$_1$. 
Similarly,   a particle reaching $0^+$ with  $v_1<0$ at $t^-$, should cross with probability  $q_{+-} = \min(1, e^{-U(0^-)+ U(0^+)  })$.
Note that this is basically a Metropolis acceptance condition, with the additional rule of inverting the velocity upon rejection.

To see that this transition rule preserves $p(\y)p(\vv)$,  
note that  a particle at  $0^-$ with  $v_1 <0$ at $t^+$ can only be the result of either 
i) a particle arrived at $0^-$ with  $v_1 > 0$, tried to cross with probability $q_{-+}$ and was rejected, or ii) a particle arrived at $0^+$ with  $v_1 < 0$, and crossed successfully to $0^-$ with probability $q_{+-}$ (obtained by inverting the signs in (\ref{qpm})).
Considering these two possibilities, we get 
\eqan 
p_{t^+}(0^-)p(v_1<0) &=& (1-q_{-+})p_{t^-}(0^-)p(v_1>0) + q_{+-} p_{t^-}(0^+)p(v_1<0)
\\
&=& p_{t^-}(0^-)p(v_1<0)
\enan 
since $p(v_1>0)=p(v_1<0)$ and $q_{-+}p_{t^-}(0^-) = q_{+-} p_{t^-}(0^+)$, and thus the probability is preserved.
Note that this last equation is the detailed balance condition, 
although the BPS sampler at continuous points does not satisfy detailed balance.

The BPS algorithm, using the inverse CDF method, can be generalized to include such discontinuities in $U(\y)$. For this we define a piecewise continuous CDF 
\eqan 
w_{\vv}(t)=1- e^{-\int_{0}^{t_1^-} \! dt' [\vv \cdot \nabla U(y_0 + \vv t')]_+ } q^1_{-+} e^{-\int_{t_1^+}^{t_2^-} \! dt' [\vv \cdot \nabla U(y_0 + \vv t')]_+ } \ldots  
q^n_{-+} e^{-\int_{t_n^+}^t \! dt' [\vv \cdot \nabla U(y_0 + \vv t')]_+ }  \,,
\enan  
with discontinuities at those times $t_i$ where the particle encounters a positive gap at $U(\y)$.
The algorithm now initializes $(\y_0, \vv)$ and then 
iterates  over the following steps:

\begin{enumerate}
 \item  Sample a uniform number $u \in [0,1]$
 
 \item Find 

 \eqan 
  t =  \sup \{ t' \, | \,\ w_{\vv}(t') \leq u 
  \}  
 \enan 
 

and move $\y$ in a straight line,
\eqan 
\y_{k+1} = \y_{k} + \vv t \,.
\enan 
\item If $\y_{k+1}$ is at a differentiable point, reflect the velocity as in (\ref{vvr}).  
Otherwise, $t=t_{i}^-$ for some~$i$, reflect $\vv$ with respect to the discontinuity plane.
\end{enumerate}

\section{Binary Distributions}
We consider now a distribution $p(\s)$ over binary variables $\s \in \{\pm 1\}^d$. 
Such a distribution can be mapped into a piecewise differentiable distribution 
using the method of~\cite{pakman2013auxiliary}, which we summarize here.
The idea is to 
augment the distribution $p(\s)$ with continuous variables $\y \in \mathbb{R}^d$ distributed as 
\eqan 
p_G(\y|\s) = 
\left\{
 \begin{array}{ll}
(2/\pi)^{d/2}\, e^{- \frac{\y \cdot \y}{2}}   &  \textrm{for} \,\,  \textrm{sign}(y_i)=s_i, \qquad  i =1, \ldots, d
\\
0 & \textrm{otherwise} \,,
\end{array}
\right.  
\label{pys}
\enan
or 
\eqan 
p_E(\y|\s) = 
\left\{
 \begin{array}{ll}
e^{- |\y|}   &  \textrm{for} \,\,  \textrm{sign}(y_i)=s_i, \qquad  i =1, \ldots, d
\\
0 & \textrm{otherwise} \,,
\end{array}
\right.  
\enan
where $  |\y| = \sum_{i=1}^d |y_i| $. 
Considering first $p_G$,  the joint distribution is now 
\eqan 
p_G(\s, \y) = p(\s) p_G(\y|\s)
\enan 
We can easily marginalize over $\s$ and obtain 
\eqan 
p_G(\y) &=& \sum_{\s'} p(\s') p_G(\y|\s') 
\\
&\varpropto& e^{-\frac{\y \cdot \y}{2}} p(\s_{\y}) 
\enan 
where 
\eqan 
\s_{\y}= \textrm{sign}(\y) \,.
\label{sy}
\enan 
Note that $p_G(\y)$  is  piecewise continuous, and defines a potential energy
\eqan 
U(\y ) &=& -\log p_G(\y)
\\
&=& \frac{\y \cdot \y}{2}  -\log p(\s_{\y})  + const.
\label{Uyb}
\enan 
Using $p_E(\y)$ we obtain similarly
\eqan 
U(\y ) &=& -\log p_E(\y)
\\
&=&  |\y|  -\log p(\s_{\y})  + const.
\enan 
In order to sample from the original distribution $p(\s)$,  we  sample from either $p_G(\y)$ or $p_E(\y)$ using the method of Section 2,
and read out the values of $\s$ from (\ref{sy}). 

Other distributions where this method could be applied  are mixed binary and (truncated) Gaussian variables (such as the spike-and-slab regression~\cite{pakman2013auxiliary})
or Bayesian Lasso models~\cite{pakman2012exact}.

\begin{figure}[t!]
\begin{center}
\includegraphics[scale=.50]{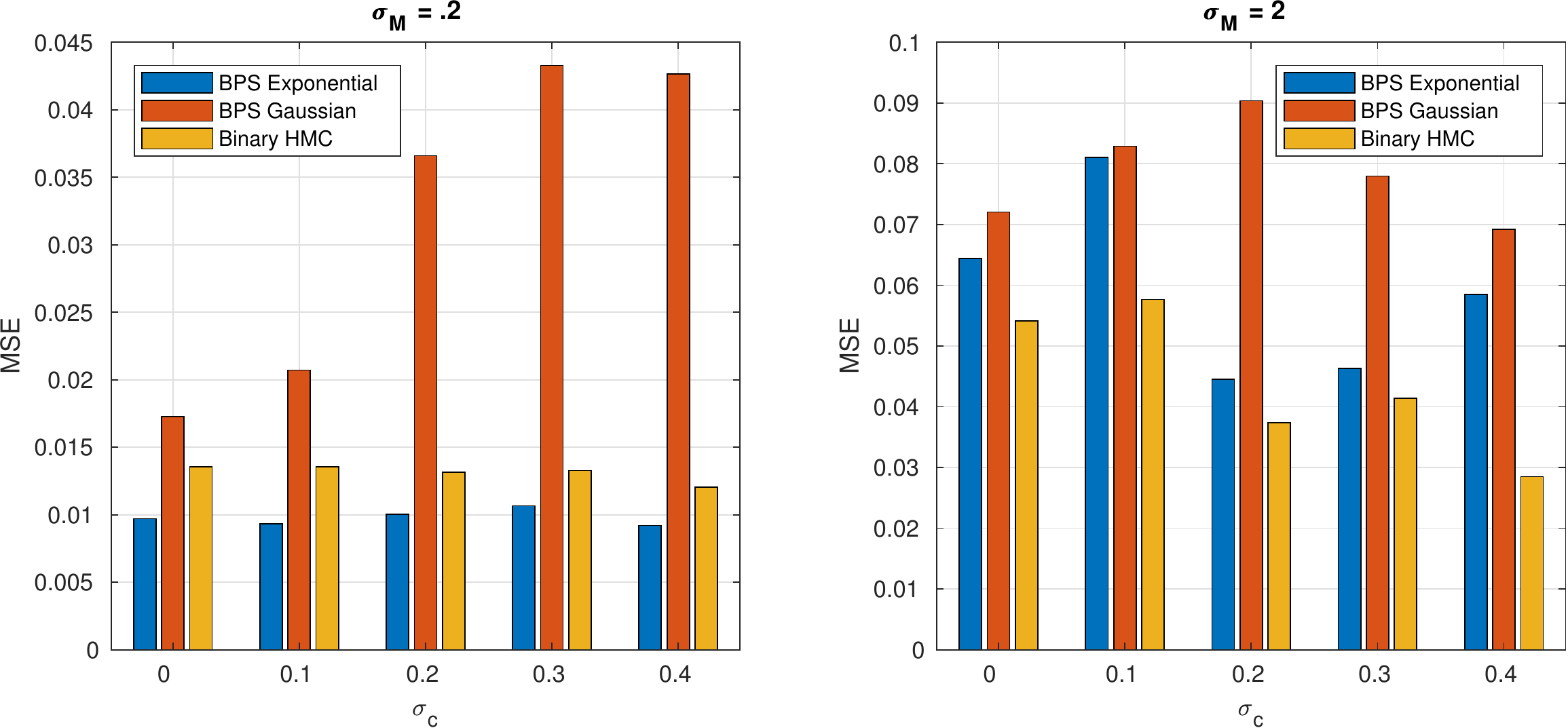}
\end{center}
\caption{MSEs of $E[s_i]$ and $E[s_i,s_j]$  for $d=10$ and different  values of the standard deviations 
$\sigma_{\M}$  and $\sigma_{{\bf r}}$  of the coefficients  in (\ref{MRF}).
The bars show the median of 30 runs, with the same CPU time for all samplers. 
The HMC travel time was $T=6.5 \pi$, but the results are similar for other~$T$s.
In this low~$d$ regime, BPS with {\it exponential} augmentation dominates for easy to mix, low $\sigma_\M$  cases. 
}
\label{MSE}
\end{figure}

\section{Example: Binary Markov Random Field}
We consider distributions of the form
\eqan 
\log p(\s) = - \s^T {\bf r}   -\frac12 \s^T \M \s \qquad \s \in \{ \pm1\}^d 
\label{MRF}
\enan 
The coefficients of $\M$ and ${\bf r}$ were sampled from  zero-mean normal distributions with standard deviations 
$\sigma_{\M}$  and $\sigma_{{\bf r}}$. The value of $\sigma_{\M}$ affects the heights of the different modes and thus controls the difficulty 
of mixing of an MCMC sampler.

We compare the binary BPS sampler, with exponential and Gaussian augmentations, with  binary HMC  with Gaussian augmentation~\cite{pakman2013auxiliary}.
All algorithms were implemented in C++ with MATLAB wrappers.\footnote{Code available at https://github.com/aripakman/binary\_bps.}

\begin{wrapfigure}{r}{7cm}
\caption{MSEs of the $E[s_i]$  for $d=100$, ${\bf r}=0$,  and different  values of the standard deviations 
$\sigma_{\M}$ of the coefficients of $\M$ in (\ref{MRF}).
The bars show the median of 30 runs, and the travel time of each HMC iteration was tuned to $T=.5 \pi$.
In this high $d$ regime, BPS with {\it Gaussian} augmentation dominates for easy to mix, low $\sigma_\M$  cases. 
}\label{MSE2}
\includegraphics[scale=.45]{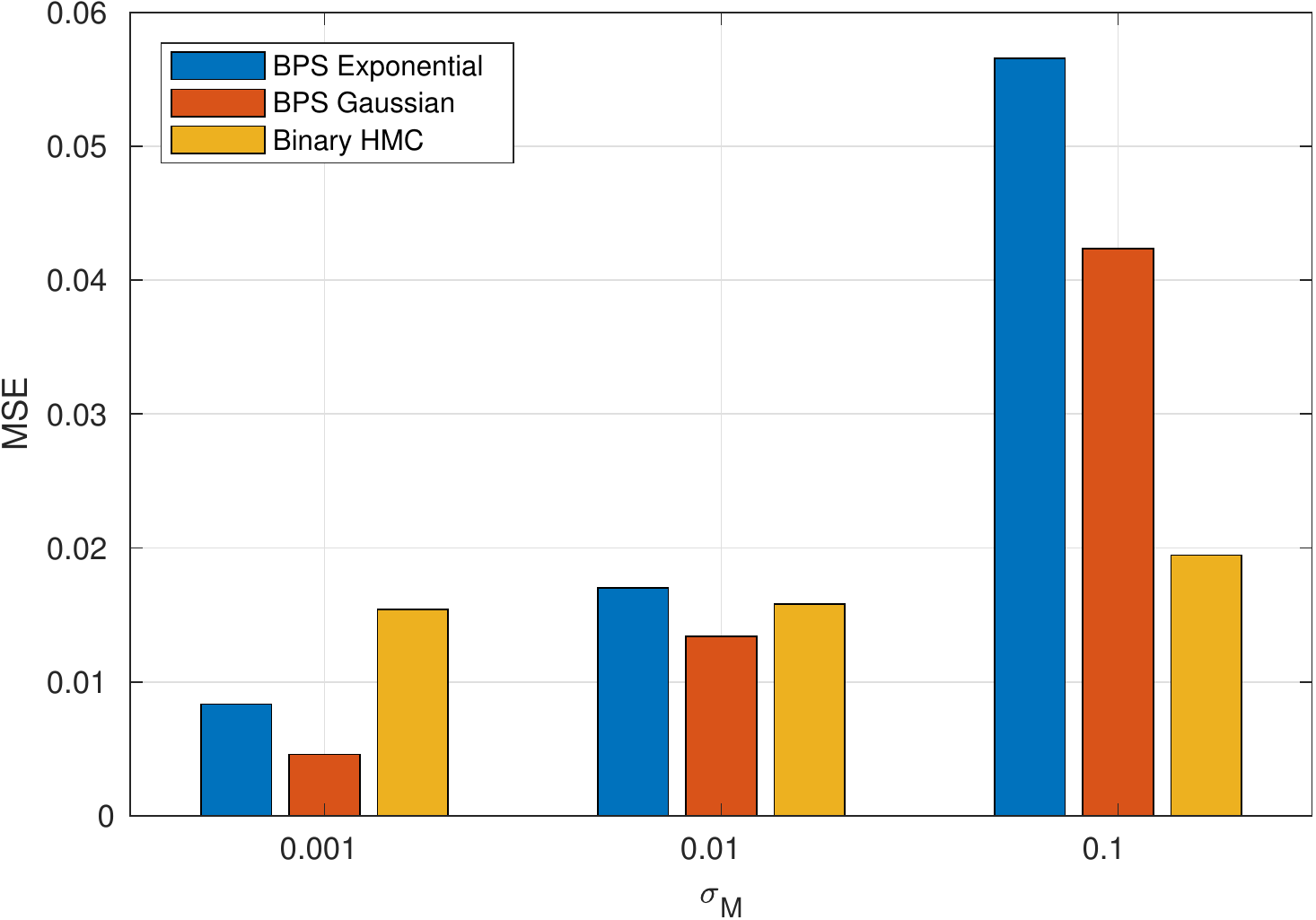}
\end{wrapfigure} 

Figure~\ref{MSE} shows the sum of the MSEs of the $E[s_i]$ and $E[s_i,s_j]$ 
for $d=10$, in easy ($\sigma_{\M}=.2$) and difficult ($\sigma_{\M}=2$) to mix regimes, 
and  different  values of $\sigma_{{\bf r}}$. 

For a fair comparison, all the samplers were run for the same CPU time. 
The results in Figure~\ref{MSE} show that  BPS with {\it exponential} augmentation is the best of the three samplers for 
easy to mix cases ($\sigma_{\M}=.2$), while  HMC  is better for the more challenging  distributions ($\sigma_{\M}=2$).

Figure~\ref{MSE2} considers the case $d=100$ and ${\bf r}=0$. In this high dimensional case, the best sampler for low $\sigma_M$ is 
BPS with {\it Gaussian} augmentation, while binary HMC dominates again in the difficult, high $\sigma_\M$ regime.

To summarize, our results show that the binary BPS samplers dominate over binary HMC  for easy to mix distributions. 
The preferred augmentation depends on the dimension:  exponential for low $d$, Gaussian for high $d$. 
This stands  in contrast to binary HMC, where the best results are obtained uniformly with the Gaussian 
augmentation~\cite{pakman2013auxiliary}.

\newpage 
\bibliographystyle{unsrt}
\bibliography{thebib}

\end{document}